# Uncovering Edge States and Electrical Inhomogeneity in MoS$_2$ Field Effect Transistors

Di Wu[1], Xiao Li[1], Lan Luan[1], Xiaoyu Wu[1], Wei Li[2], Maruthi N. Yogeesh[2], Rudresh Ghosh[2], Zhaodong Chu[1], Deji Akinwande[2], Qian Niu[1], Keji Lai[1]

[1]Department of Physics, University of Texas at Austin, Austin TX 78712, USA

[2]Microelectronics Research Center, University of Texas at Austin, Austin, TX 78758, USA

**Abstract**

The understanding of various types of disorders in atomically thin transition metal dichalcogenides (TMDs), including dangling bonds at the edges, chalcogen deficiencies in the bulk, and charges in the substrate, is of fundamental importance for their applications in electronics and photonics. Because of the imperfections, electrons moving on these two-dimensional (2D) crystals experience a spatially non-uniform Coulomb environment, whose effect on the charge transport has not been microscopically studied. Here, we report the mesoscopic conductance mapping in monolayer and few-layer MoS$_2$ field-effect transistors (FETs) by microwave impedance microscopy (MIM). The spatial evolution of the insulator-to-metal transition is clearly resolved. Interestingly, as the transistors are gradually turned on, electrical conduction emerges initially at the edges before appearing in the bulk of MoS$_2$ flakes, which can be explained by our first-principles calculations. The results unambiguously confirm that the contribution of edge states to the channel conductance is significant under the threshold voltage but negligible once the bulk of the TMD device becomes conductive. Strong conductance inhomogeneity, which is associated with the fluctuations of disorder potential in the 2D sheets, is also observed in the MIM images, providing a guideline for future improvement of the device performance.

# Introduction

Electrostatic gating in the field-effect transistor (FET) configuration has played an essential role in the blooming field of semiconducting transition metal dichalcogenides (TMDs) such as $MoS_2$ and $WSe_2$ (1). The electrical control of carrier densities in these naturally formed two-dimensional (2D) sheets is crucial for the realization of many intriguing phenomena, such as the metal-insulator transition (2-6), novel spin and valley physics (7-12), and superconducting phases (13-15). In addition, the carrier modulation provides an ideal tuning parameter to study the screening effect, which is particularly important for charge transport in 2D materials that are highly susceptible to local variations of the disorder potential (2-5, 16, 17). As a result, a complete understanding of the electronic properties of TMD FETs at all length scales, i.e., from local defects in the atomic scale, to electronic inhomogeneity in the mesoscale, to device performance in the macroscale, is imperative for both fundamental research and practical applications on these fascinating materials.

Transport and most optical measurements on TMD FETs are inherently macroscopic in nature, in which the sample response is averaged over large areas. TMD films in actual devices, however, are far from electronically uniform. Due to the relatively large amount of intrinsic defects and the inevitable charged states in the substrates, mesoscopic electrical inhomogeneity is not uncommon in TMDs, leading to hopping transport and percolation transition in the devices (6, 16-19). Little is known, however, on the magnitude and characteristic length scale of such conductance fluctuations. For layered van der Waals materials, another unique feature takes place at the sample edges, where the broken crystalline symmetry and the presence of dangling bonds introduce additional electronic states to the bulk band structure. To date, edge states in TMDs are theoretically studied by first-principles calculations (20-22) and experimentally probed by scanning tunneling microscopy (STM) and spectroscopy (SPS) (20, 23), whereas their contribution to the overall charge transport has not been fully addressed. Spatially resolved conductance maps are therefore highly desirable for the understanding of electrical inhomogeneity and edge channels in TMD FETs.

In this paper, we report the microwave impedance microscopy (MIM) (24, 25) study on the nanoscale conductance distribution during the normal operation of $MoS_2$ FETs. The experimental setup for our simultaneous transport and MIM measurements is schematically illustrated in Fig. 1*A*. The $MoS_2$ FETs and the MIM tip are mounted on the sample stage and the z-scanner of a commercial atomic-force microscope (AFM), respectively. During the contact-mode AFM scans, a low-power (~ 10 µW) 1 GHz microwave signal is delivered to the shielded cantilever tip (26) through the impedance match section. A dc bias can also be coupled to the tip using a bias-tee. The reflected signal is then amplified and demodulated to form the MIM-Im and MIM-Re signals, which are proportional to the imaginary and real parts of the small changes of tip-sample admittance in the measurement. Using the standard finite-element analysis (FEA) modeling (24), the local sample conductivity can be mapped out with a spatial resolution determined by the tip diameter (on the order of 100 nm) rather than the free-space wavelength ($\lambda$ = 30 cm) of the 1 GHz microwave. The conductance fluctuation in this mesoscopic regime is particularly important for macroscopic device performance.

**Results and Discussions**

The starting materials of our FETs are few-layer exfoliated and monolayer chemical-vapor-deposited (CVD) $MoS_2$ flakes. The samples were transferred to or directly grown on $SiO_2$ (285 nm)/Si substrates, after which the source and drain contacts (20 nm Ag/ 30 nm Au) were formed via conventional electron beam (e-beam) lithography and deposition (27, 28). To avoid direct contact between the metallic tip and the 2D sheets, which would strongly perturb the semiconducting $MoS_2$, we covered the device by the e-beam deposition of a thin (15 nm) layer of $Al_2O_3$. As a result, the carrier density in $MoS_2$ can be either globally modulated by the heavily doped Si back gate or locally tuned by the tip as a scanning top gate. Thanks to the capacitive tip-sample interaction, the MIM is capable of preforming sub-surface electrical imaging (29-32) on the buried $MoS_2$ nano-sheets.

Multiple exfoliated and CVD $MoS_2$ devices were investigated in this work, all of which exhibited similar behaviors. Fig. 1*B* shows the optical and the close-up AFM images of a typical FET fabricated on an exfoliated flake. The sample consists of two distinct regions with thicknesses of 2.1 and 2.8 nm, corresponding to 3 and 4 monolayers (MLs) of $MoS_2$, respectively. Excluding several wrinkles from the exfoliation, the surface roughness of the sample is about 0.4 nm, presumably due to the fabrication process and $Al_2O_3$ deposition. The linear output characteristic $I_{DS}$-$V_{DS}$ curves in the inset of Fig. 1*C* at different back gate voltages ($V_{BG}$) are indicative of the good Ohmic contacts between Ag and $MoS_2$ (27). From the *n*-type transfer characteristics in Fig. 1*C*, the field effect mobility can be extracted by using the expression $\mu_{FE} = (dI_{DS}/dV_{BG}) \cdot (L/W) \cdot C_{ox}^{-1} \cdot V_{DS}^{-1} \approx 5$ cm$^2$/(V·s), where $C_{ox}$ is the parallel-plate capacitance of the $SiO_2$ layer, and *L* and *W* are the channel length and width, respectively. Note that while the room-temperature mobility is comparable to that of most back-gated devices reported in the literature (3, 5, 6, 16, 17), it is much lower than the theoretical phonon-limited value (33), suggesting the presence of considerable disorder in this device, which will be explored by the MIM study below.

Fig. 2*A* displays selected MIM images within the channel region of the device in Fig. 1*B* as a function of $V_{BG}$. The complete set of data and a video clip showing the gate dependence can be found in Supplementary Information (SI), Fig. S1. The evolution of local conductance maps vividly demonstrates the insulator-to-metal transition induced by the electrostatic field effect. When the flake is in the insulating limit at $V_{BG} = -30$ V, there is virtually no electrical contrast between MoS$_2$ and the substrate. As $V_{BG}$ gradually goes up to 0 V, the contrast first emerges at the edges of the flake and then in the interior of the sample. Note that the MIM-Im signals are always higher on the 4 ML region, with slightly smaller band gap (34) and higher mobility (35) than the 3 ML part. For increasing $V_{BG}$ toward 20 V, strong inhomogeneity is observed in both MIM output channels. At the same time, the MIM signals at the edges gradually merge into the bulk and become indistinguishable with the rest of the flake for $V_{BG} > 20$ V. For even higher back-gate voltages, the sample appears uniformly bright in MIM-Im and dim in MIM-Re. During the entire process, the MIM-Im signals on the MoS$_2$ flake rise monotonically as increasing $V_{BG}$, while the MIM-Re signals reach a peak at $V_{BG} \sim 20$ V and diminish afterwards. Similar MIM data are observed by gradually ramping up the dc tip bias $V_{TG}$ during the scans, as shown in SI, Fig. S2. The local gating, on the other hand, results in a complex in-plane potential gradient away from the tip (36) and therefore will not be analyzed in detail here. We emphasize that the same trend of MIM response has been seen in all 6 MoS$_2$ FETs in this study. Fig. 2*B* shows the optical, AFM, and MIM images of a CVD-grown monolayer MoS$_2$ device (complete set of data included in SI, Fig. S3), with the overall behavior similar to that in Fig. 2*A*. In the following, we will focus on the exfoliated sample in Fig. 1 for quantitative analysis of the MIM data.

The average MIM-Im/Re signals within two 4 μm × 2 μm areas (white dashed boxes in Fig. 2*A*) on the 4 ML and 3 ML segments are shown in Fig. 2*C*, where the *x*-axis is converted from $V_{BG}$ to the two-terminal source-drain conductance ($G_{DS}$) by using the transfer curve in Fig. 1*C*. In order to quantitatively interpret the MIM signals as local conductance, we have calculated the MIM response curves as a function of the bulk MoS$_2$ sheet conductance $g_{bulk}$ using finite-element analysis (FEA) (24). Details of the FEA modeling and justification of the simulation parameters are included in SI, Fig. S4. As shown in Fig. 2*D*, the simulated MIM-Im signal, which is proportional to the tip-sample capacitance, increases monotonically as increasing $g_{bulk}$ and saturates at both the insulating ($g_{bulk} < 10^{-9}$ S·sq) and conducting ($g_{bulk} > 10^{-5}$ S·sq) limits. The MIM-Re signal, on the other hand, represents the effective loss in the tip-sample interaction and peaks at an intermediate $g_{bulk}$ of ~ $10^{-7}$ S·sq. Note that $G_{DS} \approx g_{bulk} \cdot W/L$ if the source/drain contact resistance is relatively small compared to the channel resistance. An excellent agreement between the experimental data and our modeling result can be obtained by comparing the averaged MIM signals in Fig. 2*C* and the simulation in Fig. 2*D*. The MIM images thus provide a quantitative measure of the mesoscopic conductance distribution in the MoS$_2$ FET.

A prominent feature in Fig. 2*A* is the emergence of conductive edge states before the bulk of the sample is populated by conduction electrons. The presence of localized edge channels on the boundary of a 2D system is one of the most intriguing phenomena in condensed matter physics (37). In the case of TMDs, both density functional theory (DFT) calculations (20-22) and scanning tunneling microscopy (STM) measurements (20, 23) have revealed the metallic (semiconducting) states at the zigzag (armchair) edges, while their influence on the device performance has not been experimentally probed. The MIM-Im images on the 3 ML side of the sample at selected $V_{BG}$'s are shown in Fig. 3*A*, with the line profiles across the sample edge (yellow dashed line) plotted in Fig. 3*B*. Note that the apparent width of 200 ~ 300 nm at different locations of the boundary is determined by the spatial resolution or the tip diameter $d$ rather than the actual width $w_{edge}$ of the edge states. To quantify the edge conductance, we performed 3D FEA modeling of the MIM response, in which a narrow conductive channel with $w_{edge}$ = 5 nm is situated in between the insulating substrate and the MoS$_2$ bulk. Since $w_{edge} << d$, the simulation result is invariant with respect to the product of $w_{edge}$ and the sheet conductance of the edge $g_{edge}$. Details and justifications of the modeling parameters are shown in SI, Fig. S5. The edge and bulk conductance values determined by comparing the 3D FEA and the MIM data are plotted in Fig. 3*C*. As $V_{BG}$ increases from -30 V to 0 V, $g_{edge}$ rapidly increases for more than two orders of magnitude, while the bulk conductance $g_{bulk}$ stays below our sensitivity limit of ~ $10^{-9}$ S·sq. For 0 V < $V_{BG}$ < 5 V, $g_{edge}$ levels off and $g_{bulk}$ starts to rise above the noise floor. For $V_{BG}$ above 5 V, $g_{edge}$ saturates around $4 \times 10^{-5}$ S·sq, while $g_{bulk}$ continues to increase as increasing $V_{BG}$. The quantitative mapping of $g_{edge}$ signifies the fundamental difference between the topologically trivial edge states in TMDs and the nontrivial quantum Hall (QH) (37) or quantum spin Hall (QSH) (38, 39) edges. The TMD edges in our case can be treated as a normal 1D conductor, where the carriers experience the usual scattering events. Given the FET channel length $L \approx 10$ μm in this sample, the maximum contribution of the edge states to the total conduction $G_{edge} = g_{edge} \cdot w_{edge}/L$ is on the order of $10^{-8}$ S, which is negligible once the FET is turned on. The QH or QSH edge channels, on the other hand, are dissipationless due to the suppression of back scattering and are responsible for the transport quantization when the bulk is insulating. Indeed, in our previous MIM studies on both QH (30) and QSH (31) systems, the highly conductive edges display maximum MIM-Im and zero MIM-Re signals, in sharp contrast to the finite signals in both channels for the $MoS_2$ edges.

For further investigations of the observed edge states, we carried out first-principles DFT calculations (see Methods) on 1 ML and 3 ML $MoS_2$ nano-sheets. Fig. 4*A* shows the computed energy bands of an infinite 2D sheet and a nanoribbon with armchair edges of 3 ML $MoS_2$. Compared with the bulk band structure, multiple bands appear within the bulk gap for the nanoribbon, reducing its energy gap from the bulk value of ~1.1 eV to ~ 0.3 eV, as also indicated by the density of states (DOS) plots in Fig. 4*B*. The charge density of additional electronic states at the *M* point of the Brillouin zone, circled in Fig. 4*A*, is well localized at boundary atoms in Fig. 4*C*, confirming that the additional bands in Fig. 4*A* are associated with the edge states. For a nanoribbon with zigzag edges, there are also edge states within the bandgap, which connect the bulk conduction and valence bands as shown in SI, Fig. S6 (18-20). For completeness, we have also included the DFT results of 1 ML $MoS_2$ in SI, Fig. S6, showing similar characteristics to the 3 ML $MoS_2$. In reality, the edges in our devices may be a mixture of armchair and zigzag configurations and the dangling bonds are usually terminated by foreign molecules from the environment. Nevertheless, the additional DOS inside the bulk band gap will be retained, leading to the topologically trivial edge states at the sample boundary (20, 22). As schematically illustrated in Fig. 4*D*, electrons will first populate the edge states with increasing $V_{BG}$. After these in-gap states are completely filled and $g_{edge}$ saturated, the further increase of $V_{BG}$ will then raise the Fermi level $E_F$ into the bulk conduction band, resulting in the upturn of $g_{bulk}$. Compared with the bulk bands, the edge bands are relatively flat, suggestive of higher effective mass and possibly lower mobility of the edge states. Consequently, once the bulk states participate in the transport at high gate voltages, the contribution of the edge states to the overall conductance becomes negligible. Such a physical picture nicely matches the evolution of local conductance maps deduced from the MIM data in Fig. 3c.

From the onset of bulk conduction at $V_{BG}$ = 5 V to the saturation of MIM signals around $V_{BG}$ = 25 V, pronounced spatial inhomogeneity with sub-micrometer length scale can be observed inside the MoS$_2$ flake. Both the strengths and spatial dimensions of the conductance fluctuations in this sub-threshold regime are of critical importance for the device performance. We emphasize that the large variation of MIM signals cannot be accounted for by the surface roughness of the Al$_2$O$_3$ capping layer, as analyzed in SI, Fig. S7. In Fig. 5*A*, the zoom-in MIM-Im images are displayed at $V_{BG}$ = -20 V, 14 V, and 35 V, together with the corresponding line profiles in Fig. 5*B*. The variation of MIM signals at $V_{BG}$ = 14 V corresponds to local fluctuations of $g_{bulk}$ from $5\times10^{-8}$ to $2\times10^{-7}$ S·sq in the 4 ML region. Admittedly, the MIM-Im response (Fig. 2*C*) is also highly sensitive in this intermediate conductance range. Since the transistor is completely turned off for $V_{BG}$ = -20 V ($E_F$ below the bulk conduction band minimum $E_C$) and on for $V_{BG}$ = 35 V ($E_F$ well above $E_C$), as shown in Fig. 1*C*, it is reasonable to expect much weaker spatial conductance variations under these two conditions. On the other hand, electrical inhomogeneity is most significant when $E_F$ intersects the spatially fluctuating $E_C$, as schematically shown in Fig. 5*C*. The MIM maps thus provide both quantitative measurements and direct visualization of the mesoscopic potential landscape in the sample (32), which is the combined effect from defects within the MoS$_2$ layer, charges inside the substrate and the capping layer, and impurities across the interface. The percolation network is vividly demonstrated in the sub-threshold regime, which was only indirectly inferred from atomic scale or macroscale studies (18, 19). Note that a similar technique, the alternating current scanning tunneling microscopy (ACSTM) (40), can be applied to study the atomic-scale defects in MoS$_2$, which would provide complementary information to our MIM work.

In summary, we demonstrate the local conductance mapping of functional MoS$_2$ field-effect transistors by microwave impedance microscopy. We find that, during the insulator-to-metal transition, electrons induced by the electrostatic field effect first populate the edge states before occupying the bulk of the 2D sheets, a scenario further corroborated by our first-principles calculations. The results unambiguously confirm that the contribution of edge states to the channel conductance is significant under the threshold voltage but negligible once the bulk becomes conductive. The magnitude and spatial dimensions of mesoscopic electrical inhomogeneity in the sub-threshold regime are also visualized from the MIM data. The simultaneous macroscopic transport and mesoscopic imaging experiments on TMD FETs are critically important for both fundamental research and practical applications on these fascinating materials.

## Methods

**Transport measurements:** Output and transfer characteristic curves are measured at the ambient condition with a semiconductor analyzer Keithley 4200.

**Microwave Impedance Microscopy measurements:** The MIM in this work is based on an AFM platform (Park AFM XE-70). The customized shielded cantilevers are commercially available from PrimeNano Inc. Finite-element analysis is performed using the commercial software COMSOL 4.4.

**Density functional theory calculations:** Density-functional theory calculations are performed using the Vienna *ab initio* simulation package (VASP) with the projector augmented wave method and Perdew-Burke-Ernzerhof exchange-correlation functional (41-43). For 3 ML $MoS_2$, the interlayer van der Waals interactions are taken into account by *Grimme's* D2 correction (44). A plane-wave cutoff of 400 eV and a ***k***-point spacing smaller than 0.2 Å$^{-1}$ along each periodic direction are used. A vacuum layer of larger than 15 Å is adopted to minimize the interaction between $MoS_2$ nanoribbon/sheet and its periodic images. Atomic structures are fully relaxed with the force on each atom less than 0.01 eV/ Å. The *spin-orbit coupling is included in the calculation of electronic properties, given a giant spin-orbit induced spin splitting in $MoS_2$ layers (45).* Denser ***k*** meshes and a Gaussian smearing of 0.1 eV are applied in the calculation of density of states. For an infinite sheet of 1 ML and 3 ML $MoS_2$, we construct supercells with the similar structure to corresponding armchair nanoribbons but without edges, which can be well compared with the results of armchair nanoribbons.

## Acknowledgements

The MIM work (D.W., L.L., X.W., Z.C. and K.L.) was supported by the US Department of Energy (DOE), Office of Science, Basic Energy Sciences, under Early Career Award DE-SC0010308. D.W. also acknowledges the support from the Welch Foundation Grant F-1814. W. L., M.Y. and D.A acknowledge the support of Office of Naval Research (ONR) under contract no. N00014-1110190 and the NSF NASCENT Engineering Research Center under Cooperative Agreement No. EEC-1160494. Theoretical calculation (X. L. and Q. N.) was supported by China 973 Program (Projects 2013CB921900 and 2012CB921300), DOE (DE-FG03-02ER45958) and Welch Foundation (F-1255).

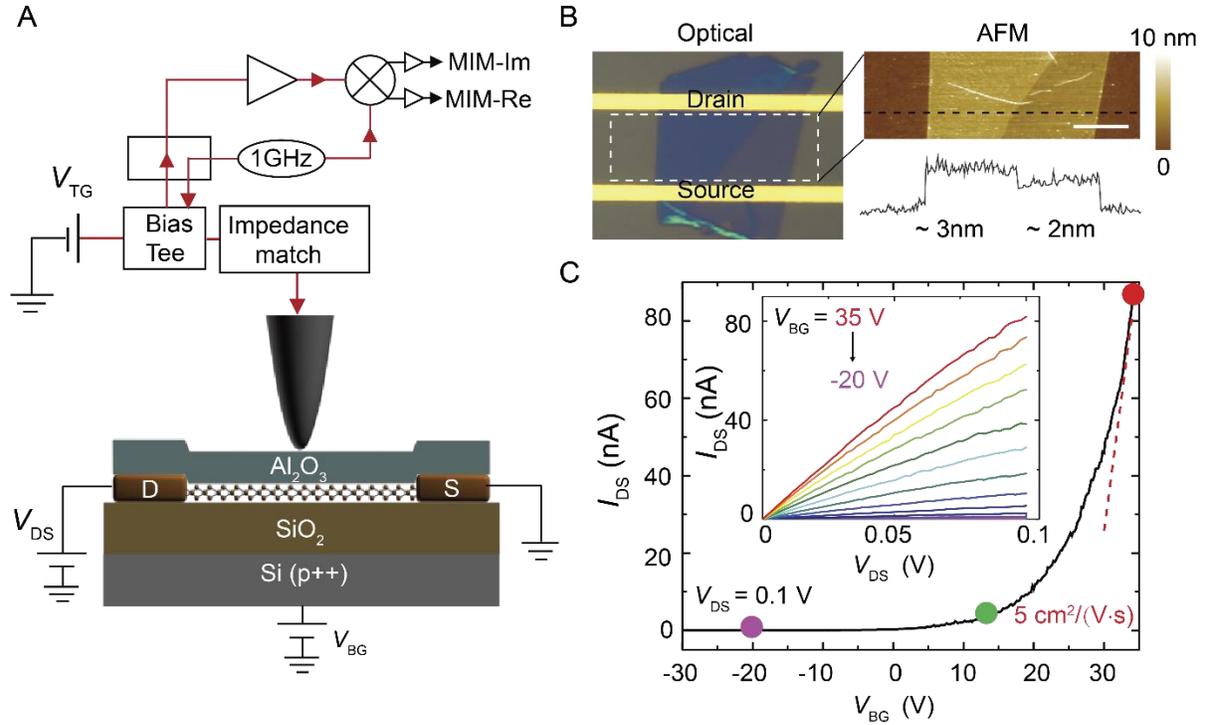

**Fig. 1.** Experimental setup and device characterization. (*A*) Schematic diagram of the device and the MIM setup. The 1 GHz microwave signal is guided to the tip through an impedance-match section, and the reflected signal is detected by the MIM electronics. The carrier density can be either globally tuned by the back-gate voltage $V_{BG}$ or locally modulated by the dc bias on the tip $V_{TG}$. (*B*) Optical and the zoom-in AFM images of an exfoliated $MoS_2$ FET device. The inset shows a line profile across the surface. The scale bar is 5 μm. (*C*) Transfer characteristics of the device at $V_{DS} = 0.1$ V. The dashed line is a linear fit to the curve for $V_{BG} > 30$ V, from which the field-effect mobility $\mu_{FE} \approx 5$ cm$^2$/(V·s) can be deduced. The solid circles (purple, green, and red) match the color coding in Fig. 5. The inset shows the output characteristics from $V_{BG} = 35$ V to -20 V with 5 V steps.

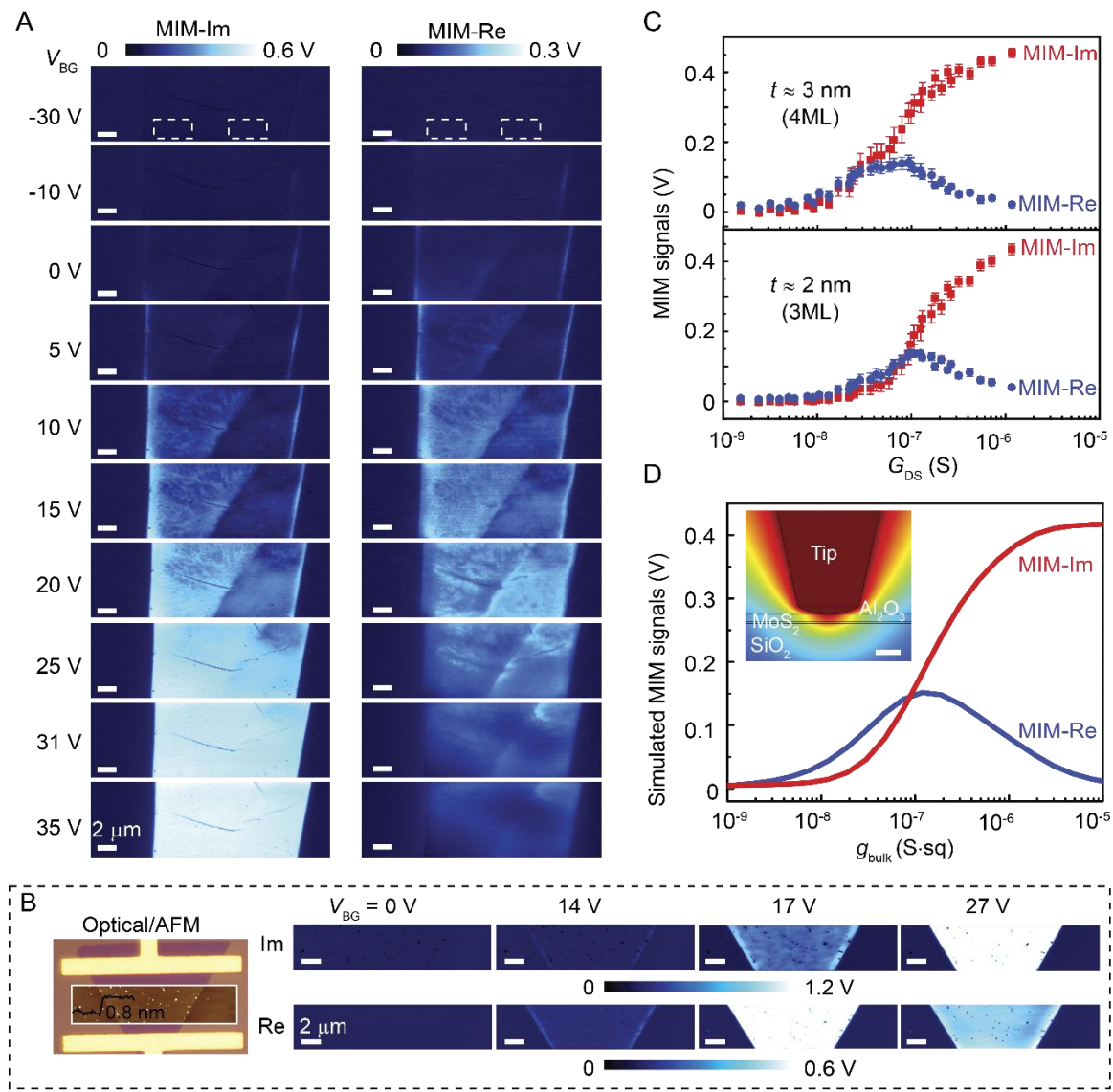

**Fig. 2.** Overall MIM response during the insulator-to-metal transition. (*A*) MIM-Im and MIM-Re images in the channel region (zoom-in image in Fig. 1B) of the device at selected back-gate voltages. (*B*) Optical, AFM (inset), and MIM images of another FET device fabricated on a CVD-grown monolayer $MoS_2$ flake (see Fig. S3 for details). All scale bars in (A) and (B) are 2 μm. (*C*) Average MIM signals inside the white dashed boxes in (A) as a function of the source-drain conductance $G_{DS}$. (*D*) Simulated MIM signals as a function of the bulk sheet conductance $g_{bulk}$, showing very good agreement with the measured data in (C). The inset shows the modeling geometry and the quasi-static potential distribution when the $MoS_2$ layer is insulating. The scale bar is 50 nm.

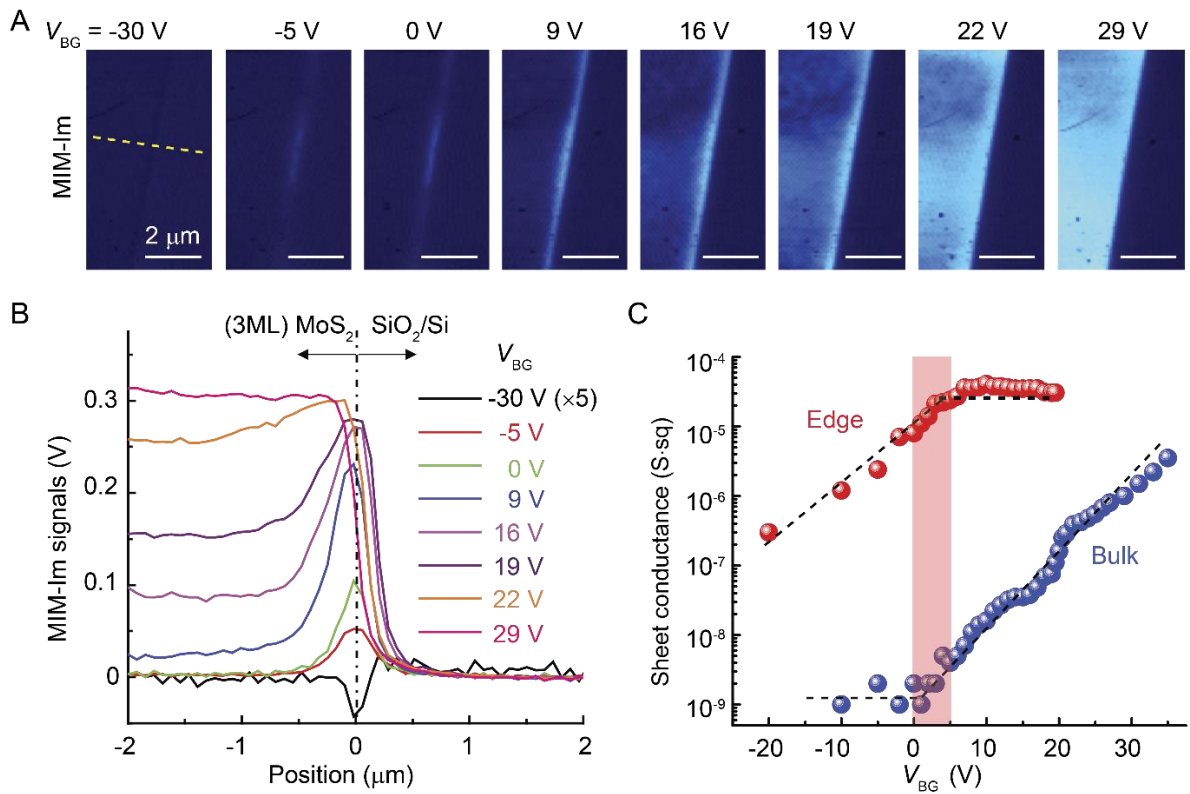

**Fig. 3.** Edge state conductance of MoS$_2$. (*A*) Selected MIM-Im images on the rightmost region of the flake. All scale bars are 2 μm. (*B*) Selected profiles (averaged over 20 lines) along the yellow dashed line in (A). The physical boundary of the 3 ML MoS$_2$ is centered in the plot. (*C*) Effective edge and bulk conductance as a function of $V_{BG}$. The shaded column marks the onset of bulk conduction and the saturation of edge conduction. The dashed lines are guides to the eyes.

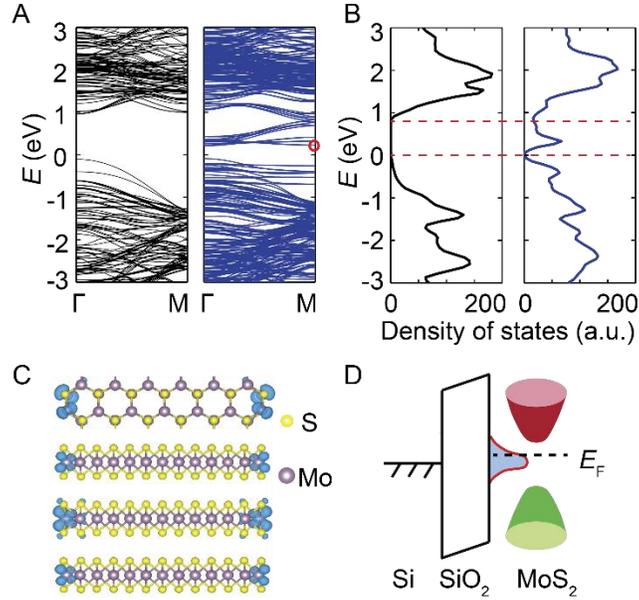

**Fig. 4.** DFT calculations of 3-layer $MoS_2$ edge states. (*A*) Calculated band structures of the 3 ML bulk $MoS_2$ (left) and a 1.9 nm-wide $MoS_2$ nanoribbon with armchair edges (right). (*B*) Total density of states of the same bulk (left) and nanoribbon $MoS_2$ (right). (*C*) Top view (top) and side view (bottom) of the electron wave functions for selected orbitals, circled in (A). The edge states are mainly localized at the boundary atoms. (*D*) Cartoon of edge and bulk band structures. As the Fermi level $E_F$ moves upwards, the edge states will be populated before the bulk states.

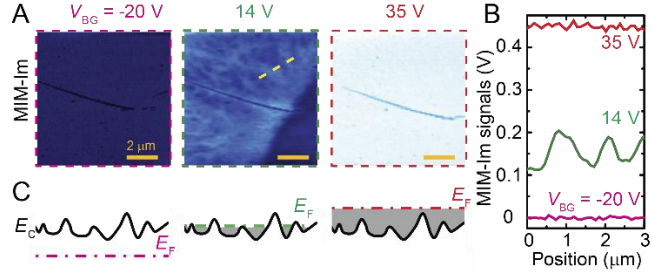

**Fig. 5.** Electrical inhomogeneity in the MoS$_2$ FET device. (*A*) Close-up MIM-Im images in the center of the flake at $V_{BG}$ = -20 V, 14 V, and 35 V. All scale bars are 2 μm. (*B*) Line cuts along the orange dashed line in (A) at the three $V_{BG}$'s. (*C*) Schematics of the relative positions between $E_F$ and the spatially fluctuating $E_C$ at the same $V_{BG}$'s above. The color coding (purple, green, and red) matches the solid circles in Fig. 1C.

# Supporting Information

# Uncovering Edge States and Electrical Inhomogeneity in MoS$_2$ Field Effect Transistors

Di Wu[1], Xiao Li[1], Lan Luan[1], Xiaoyu Wu[1], Wei Li[2], Maruthi N. Yogeesh[2], Rudresh Ghosh[2], Zhaodong Chu[1], Deji Akinwande[2], Qian Niu[1], Keji Lai[1]

[1]Department of Physics, University of Texas at Austin, Austin TX 78712, USA

[2]Microelectronics Research Center, University of Texas at Austin, Austin, TX 78758, USA

Address correspondence to kejilai@physics.utexas.edu

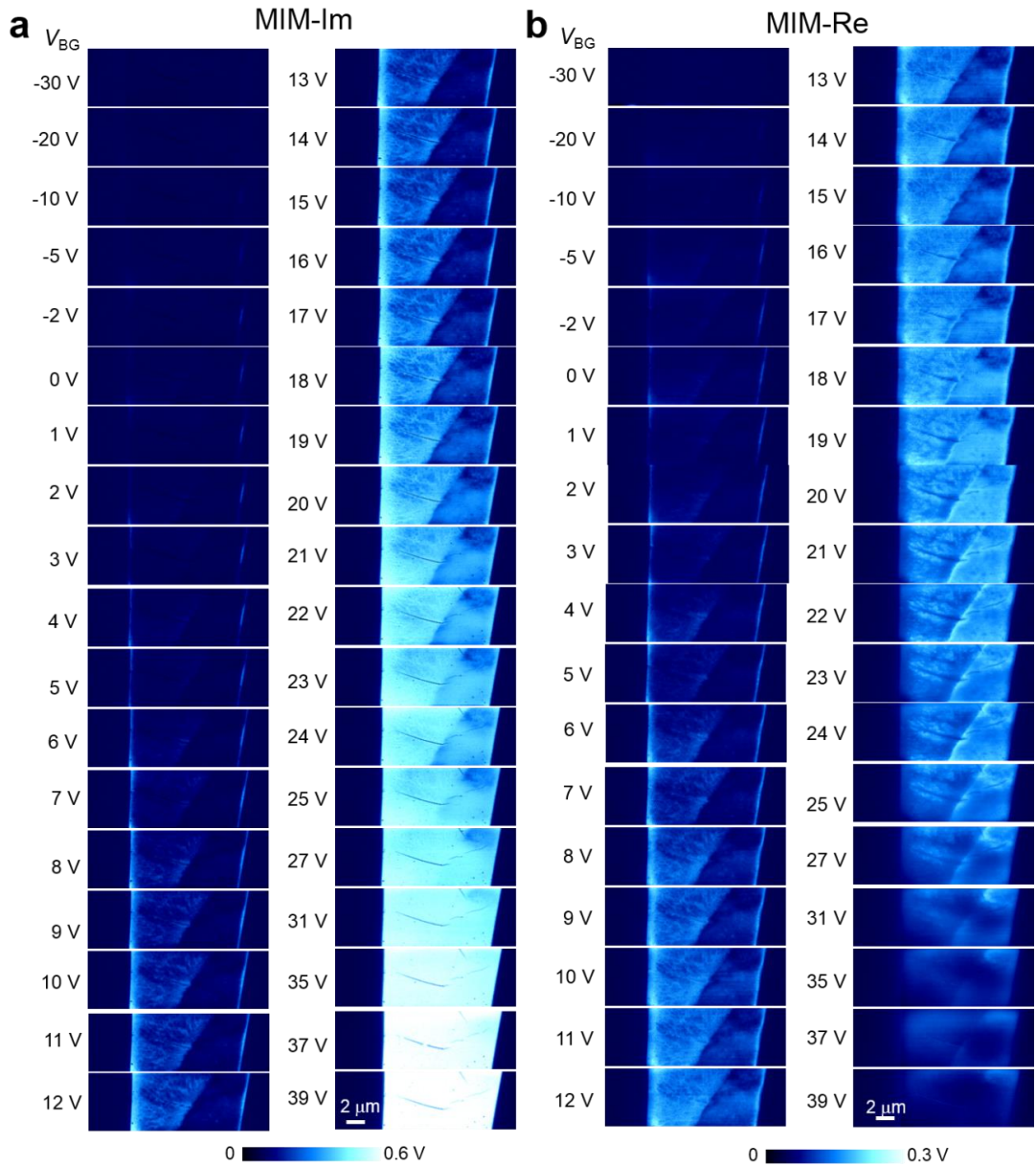

**Fig. S1.** Complete set of MIM data at different back-gate voltages. (**a**) MIM-Im and (**b**) MIM-Re images of the device channel. A video showing the evolution can also be found online.

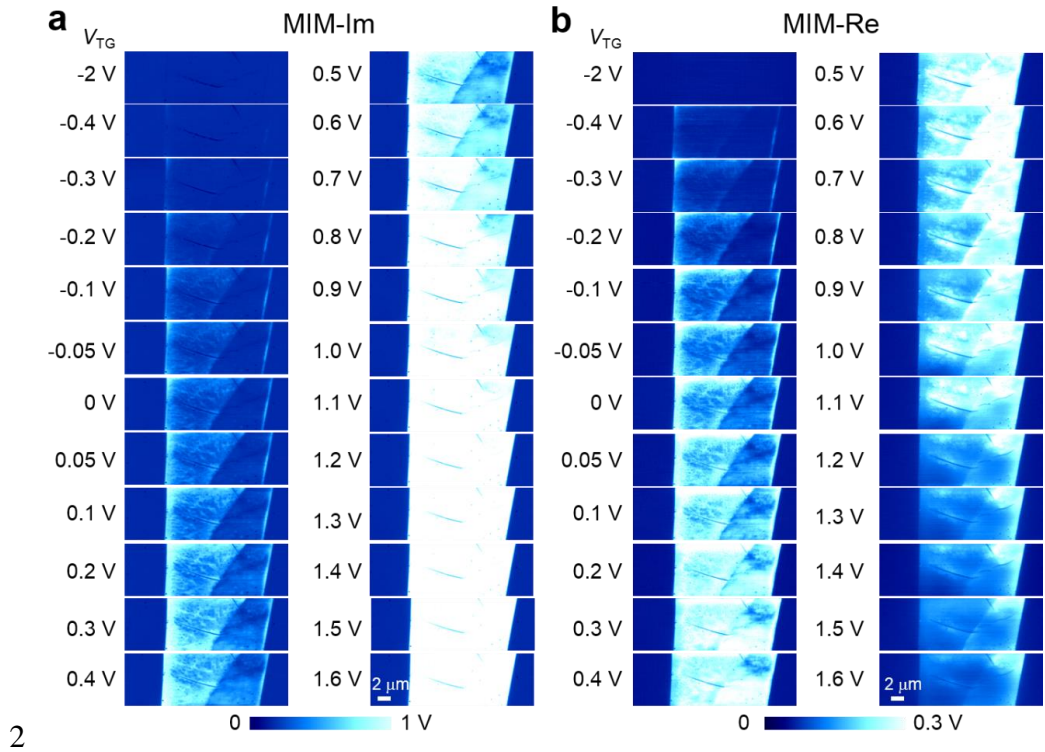

**Fig. S2.** MIM images at different top-gate voltages. (**a**) MIM-Im and (**b**) MIM-Re images on the same area of the device in Fig. 2*A*. The features are similar to that of the back-gated data. As $V_{TG}$ increases from -2 V to 1.6 V, MIM-Im signals increase monotonically while MIM-Re signals reach a peak around $V_{TG}$ = 1.5 V and then diminish. The bright edges are also observed at low $V_{TG}$'s before the bulk becomes conductive. Significant electrical inhomogeneity is visualized as well before the saturation of MIM signals.

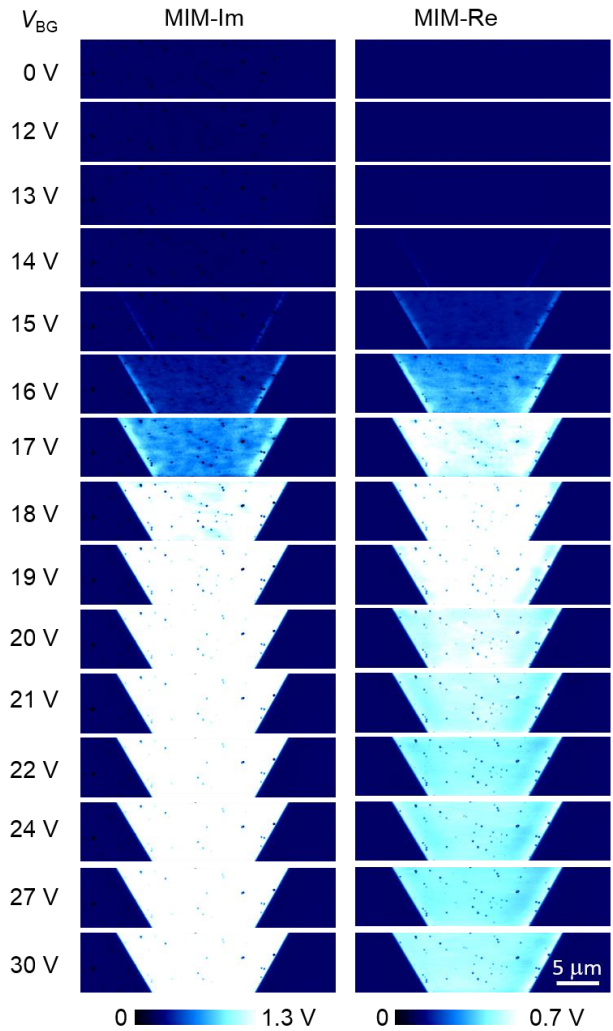

**Fig. S3.** MIM-Im (left) and MIM-Re (right) images of a device fabricated on a CVD-grown $MoS_2$ flake at different back-gate voltages. The overall evolution of MIM signals as a function of $V_{BG}$, the emergence of edge states (e.g. at $V_{BG}$ = 14 V), and the conductance inhomogeneity (e.g. at $V_{BG}$ = 17 V) are again observed in this sample. Note that different color scales compared with the data of exfoliated sample are adopted here for better visualization since a different MIM tip was used and the MIM contrasts were different. For different tips, a calibration on standard sample and Im/Re are used for quantitative analysis (46).

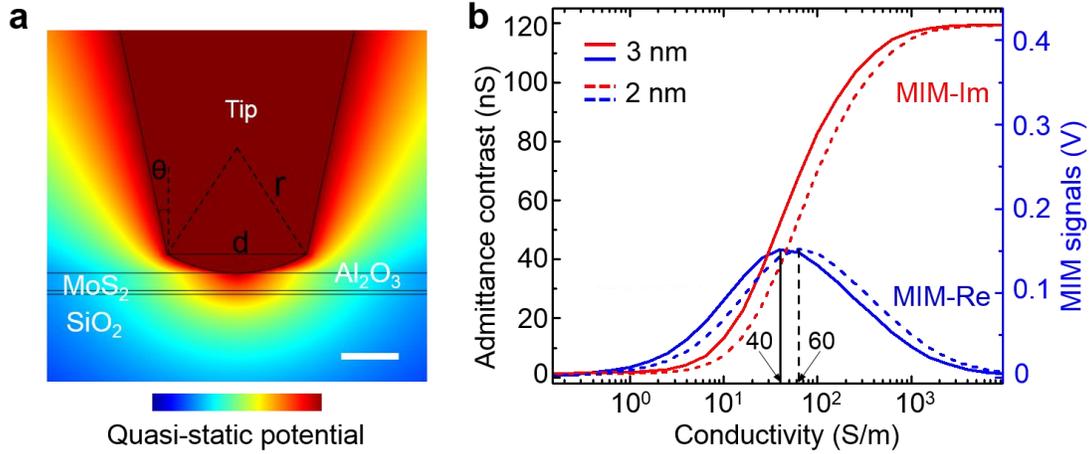

**Fig. S4.** Finite element analysis (FEA) using COMSOL 4.4 for $MoS_2$ thin films. (**a**) Modeling geometry and quasi-static potential distribution when the $MoS_2$ layer is insulating (3D conductivity $\sigma = 10^{-2}$ S/m). The scale bar is 50 nm. (**b**) Simulated admittance contrast and MIM signals as a function of $MoS_2$ conductivity $\sigma$ for both 2-nm and 3-nm-thick $MoS_2$ films.

Since all the relevant dimensions here are much smaller than the free-space wavelength (30 cm) at 1 GHz, the interaction is in the extreme near-field regime and can be modeled as lumped elements. The MIM-Im and MIM-Re signals are proportional to the imaginary and real parts of tip-sample admittance change, respectively. We have calibrated the impedance match section and the MIM electronics and shown that an admittance contrast of 1 nS corresponds to an MIM output signal of 3.5 mV (46). Since the width of the $MoS_2$ film is much larger than the MIM tip diameter, the 2D axisymmetric model can be used in this case. The shape of the MIM tip in Fig. S4a ($r = 120$ nm; $d = 120$ nm; $\theta = 12°$; tip height $h = 1$ μm) is determined by its signals on standard calibration samples (46). Other parameters are as follows, $MoS_2$: $t = 2$ nm or 3 nm in thickness, 8 μm in width and dielectric constant $\varepsilon = 7$ (6, 46, 47); $SiO_2$: 285 nm in thickness and $\varepsilon = 3.9$; heavily doped Si: conductivity $\sigma = 10^5$ S/m; $Al_2O_3$: 15 nm in thickness and $\varepsilon = 9$. The MIM response curves for the 2-nm-thick $MoS_2$ film is very similar to those of the 3-nm-thick $MoS_2$ film except that they are laterally shifted by a factor of 1.5, as shown in Fig. S4b. In other words, the MIM response is effectively invariant with respect to the 2D sheet conductance $g_{bulk} = \sigma_{bulk} \cdot t$, as long as $t$ is much smaller than the tip diameter. As a result, we only quote the sheet conductance in Fig. 2d in the main text.

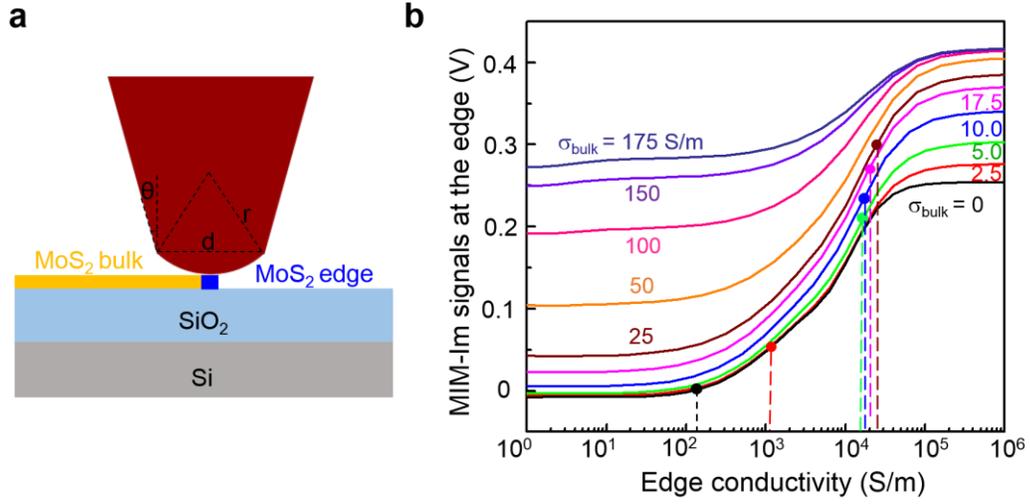

**Fig. S5.** 3D FEA modeling for MoS$_2$ with edge states. (**a**) Schematic of the tip-sample geometry of the 3D modeling. (**b**) Computed MIM-Im signals of the edge state as a function of the edge conductivity at various bulk conductivity values.

Since the width of MoS$_2$ edge states is much smaller than the tip diameter, we can no longer use the 2D axisymmetric model to simulate the edge states. Fig. S5a shows the 3D FEA geometry, in which the MoS$_2$ edge states are modeled as a thin wire with a width of $w_{edge}$ = 5 nm. Note that while the actual $w_{edge}$ can be anywhere from sub-nanometer to a few nanometers (23), its exact value is not important here because the simulation is invariant with respect to $g_{edge} \cdot w_{edge}$. As $V_{BG}$ changes, both the bulk and edge conductivity will change accordingly. Moreover, to deduce the edge conductivity, the bulk conductivity should be first determined from the 2D FEA. Figure S5b shows a series of simulated MIM-Im curves with different $\sigma_{bulk}$ from 0 to 175 S/m. The edge conductivity $\sigma_{edge}$ at a certain $V_{BG}$ can thus be extracted from the corresponding curve. The sheet conductance of edges ($g_{edge}$) in the main text is then calculated as $g_{edge} = \sigma_{edge} \cdot h$, where $h$ = 2 nm is the film thickness.

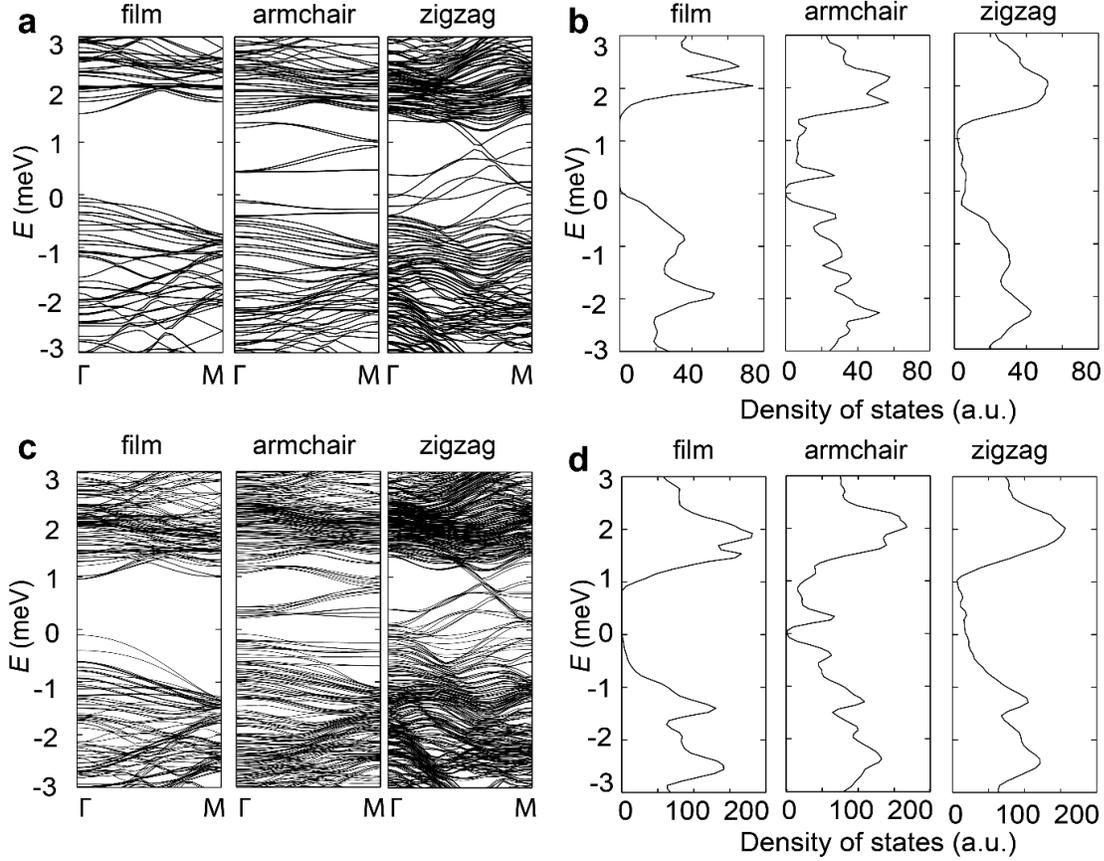

**Fig. S6.** 1 ML and 3 ML MoS$_2$ edge states by DFT calculations. (**a**) Calculated band structures of the 1 ML bulk MoS$_2$ (left), 1.9 nm-wide MoS$_2$ armchair nanoribbons (middle) and 3.1 nm-wide zigzag nanoribbons (right). For all the configurations, the edge atoms are not saturated by extra atoms, as shown in Fig. 4 in the main text. (**b**) Corresponding density of states for MoS$_2$ in (a). (**c, d**) Similar DFT results for 3 ML MoS$_2$.

Compared with bulk band structure, multiple bands emerge within the band gap for both armchair and zigzag nanoribbons. For the armchair nanoribbons, the energy gaps, ~0.5 eV for 1 ML and ~0.3 eV for 3 ML, are reduced from the bulk values, ~1.6 eV for 1 ML and ~1.1 eV for 3 ML. In contrast, the edge bands of zigzag nanoribbons seamlessly connect the bulk conduction and valence bands. Note that the spin-orbit coupling (SOC) is considered for both cases. SOC reduces the degeneracy of energy bands but does not change the main features of band structures (21).

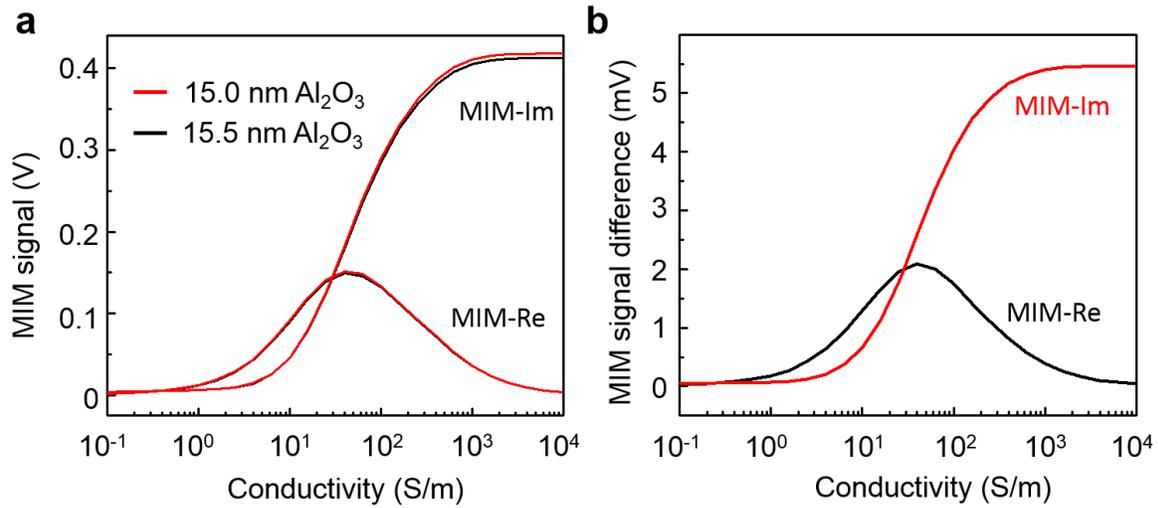

**Fig. S7.** Influence of $Al_2O_3$ roughness on MIM signals. (**a**) Computed MIM signals as a function of the $MoS_2$ conductivity for $Al_2O_3$ thicknesses of 15.0 nm and 15.5 nm, respectively. (**b**) Difference between the MIM signals for the two cases in (a). The influence of the surface roughness (0.4 ~ 0.5 nm) of the $Al_2O_3$ layer is minimal (< 6 mV) and cannot explain the large electrical inhomogeneity observed in the MIM images around the sub-threshold regime.